\documentstyle[prb,preprint,aps,epsf]{revtex}
\begin{document}
\draft
\title{
Size--dependent Correlation Effects in Ultrafast Optical Dynamics of 
Metal Nanoparticles
}
\author{T. V. Shahbazyan and I. E. Perakis} 
\address{Department of Physics and Astronomy, 
Vanderbilt University, Box 1807-B,  
Nashville, TN 37235}

\maketitle

\begin{abstract}
We study the role of collective surface excitations in the electron
relaxation in small metal particles. We show that the
dynamically screened electron--electron interaction in a nanoparticle 
contains a size--dependent correction induced by the surface. This leads
to new channels of  quasiparticle scattering accompanied by the
emission of surface collective excitations. We calculate the energy
and temperature dependence of the 
corresponding rates, which depend strongly on the nanoparticle size. 
We  show that the  surface--plasmon--mediated
scattering rate of a conduction electron increases with energy, in
contrast to that mediated by a bulk plasmon. 
In noble--metal particles, we find that the 
dipole collective excitations (surface plasmons) mediate a 
resonant scattering of $d$--holes to the conduction band.
We study the role of the latter  effect in the
ultrafast optical dynamics of small nanoparticles and show that,
with decreasing nanoparticle size,
it leads to a drastic change in the differential absorption
lineshape and  a strong frequency dependence of the relaxation near
the surface plasmon resonance. The experimental implications of our
results in ultrafast pump--probe spectroscopy are also discussed.
\end{abstract}


\section{Introduction}

The properties of small metal particles 
in the intermediate regime between 
bulk--like and molecular  behavior 
have been a subject of great interest 
recently. \cite{kreibig,heer,brack,cluster}
Even though the electronic and optical properties of nanoparticles
have been extensively studied,
the effect of confinement on electron dynamics
is much less understood.
Examples of outstanding issues include 
the role of electron--electron interactions 
in the process of cluster fragmentation, the role of surface 
lattice modes in providing additional
channels for intra-molecular energy relaxation, 
the influence of
the electron and nuclear motion on the superparamagnetic properties of
clusters, 
and the effect of confinement on the nonlinear optical properties 
and  transient response under ultrafast excitation. 
\cite{kreibig,heer,cluster}
These and other dynamical phenomena can be studied with
femtosecond nonlinear optical spectroscopy, which allows one to 
probe the time evolution of the excited states
with a resolution shorter than the energy 
relaxation or dephasing times.

Surface collective excitations play an important role in the
absorption of light by metal nanoparticles.
In large particles with sizes
comparable to the wave--length of light $\lambda$ (but smaller than the
bulk mean free path), 
the lineshape of the surface plasmon (SP) 
resonance is determined by  the electromagnetic effects.\cite{kreibig} 
In small nanoparticles with radii
$R\ll \lambda$, the absorption spectrum is governed  by quantum 
confinement effects. For example, 
the momentum non--conservation due to the confining potential 
leads to the Landau damping of the SP and to a resonance linewidth
inversely proportional to the nanoparticle size.\cite{kreibig,kaw66}  
Confinement changes also non--linear optical properties of
nanoparticles: a size--dependent  enhancement
of the third order susceptibilities, caused by the elastic 
surface scattering of single--particle excitations, has been
reported.\cite{hac86,aga88,yan94}  

Extensive experimental studies of 
the electron relaxation
in nanoparticles have  recently 
been performed using 
ultrafast pump--probe
spectroscopy.\cite{tok94,rob95,big95,ahm96,per97,nis97,kla98,sha98}
Unlike in semiconductors, the dephasing processes in metals
are very fast, and nonequilibrium
populations of optically excited 
electrons and holes are formed within 
several femtoseconds.  
These thermalize into the hot Fermi--Dirac distribution
within several hundreds of femtoseconds,
mainly due to  {\em e--e} and {\em h--h} scattering.
\cite{fan92,sun94,gro95,fly98}
Since the electron heat capacity is much smaller than that of the
lattice, a high electron temperature can be reached during 
less than 1 ps time scales, i.e., before any 
significant energy transfer to the phonon bath occurs. 
During this stage, the SP resonance was observed to undergo a
time--dependent spectral broadening.\cite{big95,per97} 
Subsequently, the electron and phonon baths
equilibrate  through the electron--phonon interactions 
over  time intervals of  a few picoseconds.
During this incoherent stage, 
the hot electron distribution can be characterized by a 
time--dependent temperature. 
Correlation effects play an important role in the latter 
regime. For example, in order to explain the differential 
absorption lineshape, it is essential to take into account the
{\em e--e} scattering of the optically--excited carriers near the Fermi
surface.\cite{big95} 
Furthermore, despite the similarities to the bulk--like behavior,
observed, e.g., in metal films, 
certain  aspects of the optical
dynamics in nanoparticles are 
significantly different.\cite{nis97,big95,sha98}
For example, experimental studies of small Cu
nanoparticles revealed that the relaxation times  of the the
pump--probe signal depend strongly on frequency: 
the relaxation was considerably slower
at the SP resonance.\cite{big95,sha98} 
This and other  observations suggest that 
collective surface excitations
play an important role in the electron dynamics
in small metal particles.

Let us recall the basic facts regarding the linear absorption by
metal nanoparticles embedded in a medium with
dielectric constant $\epsilon_m$. We will focus primarily on
noble metal particles containing several hundreds of atoms; in this
case, the confinement affects the extended electronic states  
even though the bulk lattice structure has been established.
When the particles radii are small, $R\ll \lambda$, so that
only dipole surface modes can be 
optically  excited and non--local effects can be neglected, the
optical properties of this system are 
determined by the dielectric function \cite{kreibig}   
\begin{equation}
\epsilon_{\rm col}(\omega)=
\epsilon_m+3p\epsilon_m\,
\frac{\epsilon(\omega)-\epsilon_m}{\epsilon(\omega)+2\epsilon_m},
\end{equation}
where $\epsilon(\omega)=\epsilon'(\omega)+i\epsilon''(\omega)$ is 
the dielectric function  of a metal particle
and $p\ll 1$ is the volume fraction
occupied by nanoparticles in the colloid.  Since the 
$d$--electrons play an important role in the optical properties of
noble metals, the dielectric function 
$\epsilon(\omega)$ includes also the interband contribution
$\epsilon_d(\omega)$. 
For $p\ll 1$, the absorption  coefficient of such a system is
proportional to that of a single particle and is given by\cite{kreibig}  
\begin{equation}\label{absor}
\alpha(\omega)= -9p\,\epsilon_m^{3/2}\,\frac{\omega}{c}\,
\mbox{Im} {1\over \epsilon_{s}(\omega)},
\end{equation}
where
\begin{equation}\label{epseff}\epsilon_{s}(\omega)=
\epsilon_d(\omega)-\omega_p^2/\omega(\omega+i\gamma_s)+2\epsilon_m,
\end{equation}
plays the role of an effective dielectric function of a particle in
the medium. Its zero, $\epsilon'_{s}(\omega_{s})=0$, determines the
frequency of the SP, $\omega_{s}$. In Eq.\ (\ref{epseff}),
$\omega_p$ is the bulk plasmon frequency of the conduction
electrons, and the width $\gamma_s$ characterizes the SP 
damping. 
The semiclassical result
Eqs.\  (\ref{absor}) and  (\ref{epseff}) applies to nanoparticles
with radii $R\gg q_{_{TF}}^{-1}$, where  
$q_{_{TF}}$ is the Thomas--Fermi screening wave--vector
($q_{_{TF}}^{-1}\sim 1$ {\AA} in noble metals).
In this case, the electron density deviates from its 
classical shape only within a surface layer 
occupying  a small fraction of the total volume.\cite{kre92}
Quantum mechanical corrections, arising from the discrete energy
spectrum, lead to a width $\gamma_s\sim v_{_F}/R$, where
$v_{_F}=k_{_F}/m$ is the Fermi velocity.\cite{kreibig,kaw66}
Even though 
$\gamma_s/\omega_s\sim (q_{_{TF}}R)^{-1}\ll 1$, this damping
mechanism dominates over  others, e.g., due to phonons, for sizes 
$R\lesssim 10$ nm. In small clusters, containing 
several dozens  of atoms, 
the semiclassical approximation breaks down and 
density functional  or {\em ab initio} methods should
be used.\cite{kreibig,heer,brack,cluster} 

It should be noted that, in contrast to surface collective
excitations, the {\em e--e} scattering is not  sensitive to the
nanoparticle size as long as  the condition 
$q_{_{TF}}R\gg 1$ holds.\cite{siv94}
Indeed, for such sizes, the static screening is essentially 
bulk--like. At the same time, the energy dependence of the bulk 
{\em e--e} scattering rate,\cite{pines} 
$\gamma_e\propto (E-E_F)^2$, with $E_F$ being the Fermi energy,
comes from the phase--space restriction 
due to the momentum conservation, and 
involves the exchange of typical  momenta $q\sim q_{_{TF}}$. 
If the size--induced momentum uncertainty 
$\delta q \sim R^{-1}$ is much smaller than $q_{_{TF}}$, the 
{\em e--e} scattering rate in a  nanoparticle is not 
significantly affected by the confinement.\cite{alt97}

In this paper we address the role of collective surface excitations 
in the electron relaxation in small metal particles. We show
that the dynamically screened {\em e--e} interaction contains a
correction originating from the surface collective 
modes excited by an electron in nanoparticle.
This opens up  new quasiparticle scattering channels 
mediated by surface collective modes. 
We derive the corresponding scattering rates, which depend strongly
on the nanoparticle size.
The scattering rate of a conduction electron increases with energy,
in contrast to the bulk--plasmon mediated scattering.  
In noble metal particles, we study the SP--mediated
scattering of a $d$--hole into the conduction band. 
The scattering rate of this process depends strongly on temperature, 
and exhibits a {\em peak} as a function of energy due to the
restricted phase space available for interband scattering. 
We show that this effect manifests itself in the 
ultrafast nonlinear optical dynamics of nanometer--sized particles.
In particular, our self--consistent calculations show
that, near the SP resonance, the differential absorption lineshape
undergoes a dramatic transformation as the particle size
decreases. We also find 
that the relaxation times of the pump--probe signal depend strongly
on the probe frequency, in agreement with recent experiments.

The paper is organized as follows. In Section \ref{sec:screen} we
derive the dynamically screened  Coulomb potential in a
nanoparticle. In Section \ref{sec:decay} we calculate the
SP--mediated quasiparticle scattering rates 
of the  conduction electrons and 
the {\em d}--band holes. 
In Section \ref{sec:optics} we incorporate these effects
in the calculation of the absorption spectrum and study their
role in the size and frequency dependence of the time--resolved
pump--probe signal.

\section{Electron--electron interactions in metal nanoparticles}
\label{sec:screen}

In this section, we study the effect of the surface collective
excitations on the {\em e--e} interactions in a spherical metal particle. 
To find the dynamically screened Coulomb potential, we
generalize the method previously developed for calculations of
local field corrections to the optical fields.\cite{lus74}
The potential 
$U(\omega;{\bf r},{\bf r}')$ at point ${\bf r}$ arising from 
an electron at point ${\bf r}'$ is determined by the equation\cite{mahan}
\begin{eqnarray}\label{dyson}
U(\omega;{\bf r},{\bf r}')=u({\bf r}-{\bf r}')+
\int d{\bf r}_1 d{\bf r}_2u({\bf r}-{\bf r}_1)
\Pi(\omega;{\bf r}_1,{\bf r}_2)U(\omega;{\bf r}_2,{\bf r}'),
\end{eqnarray}
where $u({\bf r}-{\bf r}')=e^2|{\bf r}-{\bf r}'|^{-1}$ is the unscreened
Coulomb potential
and $\Pi(\omega;{\bf r}_1,{\bf r}_2)$ is the polarization
operator.
There are three contributions to $\Pi$, 
arising from the polarization of the 
conduction electrons, the $d$--electrons, and the  medium
surrounding the nanoparticles: 
$\Pi=\Pi_c+\Pi_d+\Pi_m$. It is useful to rewrite 
Eq.\ (\ref{dyson}) in the ``classical'' form 
\begin{equation}
\label{gauss}
\nabla\cdot({\bf E}+4\pi{\bf P})=4\pi e^2\delta({\bf r}-{\bf r}'),
\end{equation}
where ${\bf E}(\omega;{\bf r},{\bf r}')=
-\nabla U(\omega;{\bf r},{\bf r}')$ is the screened Coulomb
field
and 
${\bf P}={\bf P}_c+{\bf P}_d+{\bf P}_m$ is the electric
polarization vector, related to the potential $U$ as
\begin{equation}\label{delp}
\nabla {\bf P}(\omega;{\bf r},{\bf r}')
=-e^2\int d{\bf r}_1\Pi(\omega;{\bf r},{\bf r}_1)
U(\omega;{\bf r}_1,{\bf r}').
\end{equation}
In the random phase approximation,
the intraband polarization operator is given by  
\begin{eqnarray}\label{pol}
\Pi_c(\omega;{\bf r},{\bf r}')=\sum_{\alpha\alpha'}
\frac{f(E_{\alpha}^c)-f(E_{\alpha'}^c)}
{E^c_{\alpha}-E^c_{\alpha'}+\omega+i0}
\psi^c_{\alpha}({\bf r})\psi^{c\ast}_{\alpha'}({\bf r})
\psi_{\alpha}^{c\ast}({\bf r}')\psi^c_{\alpha'}({\bf r}'),
\end{eqnarray}
where $E^c_{\alpha}$ and  $\psi^c_{\alpha}$ are the
single--electron
eigenenergies and eigenfunctions in the nanoparticle,
and $f(E)$ is the Fermi--Dirac 
 distribution (we
set $\hbar=1$). Since we are interested in
frequencies much larger than the single--particle level spacing, 
$\Pi_c(\omega)$ can be expanded in terms of
$1/\omega$. For the real part, 
$\Pi'_c(\omega)$, we obtain in the leading order\cite{lus74}
\begin{equation}\label{pol1}
\Pi'_c(\omega;{\bf r},{\bf r}_1)=
-\frac{1}{m\omega^2}\nabla [n_c({\bf r})\nabla \delta({\bf r}-{\bf r}_1)],
\end{equation}
where $n_c({\bf r})$ is the conduction electron density. In the
following we assume, for simplicity, a step density  
profile,
$n_{c}({\bf r})=\bar{n}_{c}\,\theta(R-r)$, where 
$\bar{n}_{c}$ is the average density. The leading contribution to the
imaginary part, $\Pi''_c(\omega)$,  is proportional to
$\omega^{-3}$, 
so that   $\Pi''_c(\omega)\ll \Pi'_c(\omega)$.

By using Eqs.\ (\ref{pol1}) and (\ref{delp}), one obtains a familiar
expression for 
${\bf P}_c$ at high frequencies,
\begin{equation}\label{vecc}
{\bf P}_c({\omega};{\bf r},{\bf r}')
=\frac{e^2n_c({\bf r})}{m\omega^2}\nabla U(\omega;{\bf r},{\bf r}')
=\theta(R-r)\chi_c(\omega){\bf E}(\omega;{\bf r},{\bf r}'),
\end{equation} 
where $\chi_c(\omega)=-e^2\bar{n}_c/m\omega^2$ is the conduction
electron susceptibility. Note that, for a step density  
profile, ${\bf P}_c$ vanishes
outside the particle. The  $d$--band and dielectric
medium contributions to
${\bf P}$ are also given by similar relations,
\begin{eqnarray}
{\bf P}_d({\omega};{\bf r},{\bf r}')
=\theta(R-r)\chi_d(\omega)
{\bf E}(\omega;{\bf r},{\bf r}'),\label{vecd}
\\
{\bf P}_m({\omega};{\bf r},{\bf r}')
=\theta(r-R)\chi_m
{\bf E}(\omega;{\bf r},{\bf r}'),\label{vecm}
\end{eqnarray}
where $\chi_i=(\epsilon_i-1)/4\pi$, $i=d,m$ are the corresponding
susceptibilities and  the step functions account for the boundary
conditions.\cite{boundary} Using Eqs.\ (\ref{vecc})--(\ref{vecm}),
one can write a closed equation for $U(\omega;{\bf r},{\bf r}')$. 
Using Eq.\ (\ref{delp}), 
the second term of Eq.\ (\ref{dyson}) can be presented as
%
$
-e^{-2}\int d{\bf r}_1 u({\bf r}-{\bf r}_1)
\nabla \cdot {\bf P}(\omega;{\bf r}_1,{\bf r}').
$
%
Substituting the above expressions for ${\bf P}$, 
we then obtain after 
integrating by parts
\begin{eqnarray}\label{self1}
\epsilon(\omega)
U(\omega;{\bf r},{\bf r}')=
\frac{e^2}{|{\bf r}-{\bf r}'|}
&&
+\int d{\bf r}_1 
\nabla_1\frac{1}{|{\bf r}-{\bf r}_1|}\cdot\nabla_1
\left[\theta(R-r)\chi(\omega)+\theta(r-R)\chi_m\right]
U(\omega;{\bf r}_1,{\bf r}')
\nonumber\\ &&
+
i\int d{\bf r}_1 d{\bf r}_2\frac{e^2}{|{\bf r}-{\bf r}_1|}
\Pi''_c(\omega;{\bf r}_1,{\bf r}_2)U(\omega;{\bf r}_2,{\bf r}'),
\end{eqnarray}
with
\begin{equation}
\label{eps}
\epsilon(\omega)\equiv 1+4\pi\chi(\omega)
=\epsilon_d(\omega)-\omega_p^2/\omega^2,
\end{equation}
$\omega_p^2=4\pi e^2\bar{n}_c/m$ being the
plasmon frequency in the conduction band.
The last term in the rhs of Eq.\ (\ref{self1}), proportional to
$\Pi''_c(\omega)$, can be regarded as a small correction.
To solve Eq.\ (\ref{self1}), we first eliminate the angular
dependence by expanding 
$U(\omega;{\bf r},{\bf r}')$ in spherical harmonics,
$Y_{LM}(\hat{\bf r})$, with coefficients $U_{LM}(\omega;r,r')$.
Using the corresponding expansion of 
$|{\bf r}-{\bf r}'|^{-1}$ with  coefficients 
$Q_{LM}(r,r')=\frac{4\pi}{2L+1}r^{-L-1}r'^{L}$ (for $r>r'$), 
we get the following equation for $U_{LM}(\omega;r,r')$:
\begin{eqnarray}\label{self2}
\epsilon(\omega)U_{LM}(\omega;r,r')=
&&
Q_{LM}(r,r')+
4\pi\left[\chi(\omega)-\chi_m\right]
\frac{L+1}{2L+1}\left({r\over R}\right)^L U_{LM}(\omega;R,r')
\nonumber\\ &&
+ie^2\sum_{L'M'}\int dr_1dr_2r_1^2r_2^2Q_{LM}(r,r_1)
\Pi''_{LM,L'M'}(\omega;r_1,r_2)U_{L'M'}(\omega;r_2,r'),
\end{eqnarray}
where 
\begin{eqnarray}
\Pi''_{LM,L'M'}(\omega;r_1,r_2)=
\int d\hat{\bf r}_1d\hat{\bf r}_2 
Y_{LM}^{\ast}(\hat{\bf r}_1)
\Pi''_c(\omega;{\bf r}_1,{\bf r}_2)Y_{L'M'}(\hat{\bf r}_2),
\end{eqnarray}
are the coefficients of the multipole expansion of
$\Pi''_c(\omega;{\bf r}_1,{\bf r}_2)$. For $\Pi''_c=0$, 
the solution of 
Eq.\ (\ref{self2}) can be presented in the form
\begin{eqnarray}\label{sol}
U_{LM}(\omega;r,r')=a(\omega)e^2Q_{LM}(r,r')
+b(\omega)\frac{4\pi e^2}{2L+1}\frac{r^Lr'^L}{R^{2L+1}},
\end{eqnarray}
with frequency--dependent coefficients $a$ and $b$.
Since  $\Pi''_c(\omega)\ll\Pi'_c(\omega)$ for relevant frequencies,
the solution of Eq.\ (\ref{self2}) in the
presence of the last term can be written in the
same form as Eq.\ (\ref{sol}),  but with modified $a(\omega)$ and
$b(\omega)$. Substituting  
Eq.\ (\ref{sol}) into Eq.\ (\ref{self2}), we obtain after lengthy
algebra in the lowest order in $\Pi''_c$
\begin{equation}
\label{ab}
a(\omega)=\epsilon^{-1}(\omega),
~~b(\omega)=\epsilon^{-1}_L(\omega)-\epsilon^{-1}(\omega),
\end{equation}
where
\begin{equation}\label{epsL}
\epsilon_L(\omega)={L\over 2L+1} 
\epsilon(\omega)
+{L+1\over 2L+1}\epsilon_m+i\epsilon''_{cL}(\omega),
\end{equation}
is the effective dielectric function, whose zero,
$\epsilon'_L(\omega_L)=0$, determines the frequency of the collective
surface excitation with angular momentum $L$,\cite{kreibig}
\begin{equation}
\label{omegaL}
\omega_L^2=\frac{L\omega_p^2}{L\epsilon'_d(\omega_L)+(L+1)\epsilon_m}.
\end{equation}
In Eq.\ (\ref{epsL}), $\epsilon''_{cL}(\omega)$ 
characterizes the damping of the $L$--pole collective mode by
single--particle excitations, and is given by
\begin{equation}\label{epscl}
\epsilon''_{cL}(\omega)=\frac{4\pi^2 e^2}{(2L+1)R^{2L+1}}
\sum_{\alpha\alpha'}|M^{LM}_{\alpha\alpha'}|^2
[f(E_{\alpha}^c)-f(E_{\alpha'}^c)]\delta(E^c_{\alpha}-E^c_{\alpha'}+\omega),
\end{equation}
where $M^{LM}_{\alpha\alpha'}$ are the matrix elements of 
$r^LY_{LM}(\hat{\bf r})$. 
Due to the momentum nonconservation in a nanoparticle, the matrix
elements are finite, which leads to the size--dependent width of
the $L$--pole mode:\cite{kaw66,lus74}  
\begin{equation}
\gamma_L=\frac{2L+1}{L}\frac{\omega^3}{\omega_p^2}\epsilon''_{cL}(\omega).
\end{equation}
For $\omega\sim\omega_L$, one can show that the width, 
$\gamma_L\sim v_F/R$, is independent of $\omega$.
Note that, in noble metal particles,
there is an additional {\em d}--electron contribution to 
the imaginary part of $\epsilon_L(\omega)$ at 
frequencies above the onset $\Delta$ of  the interband transitions.

Putting everything together, we arrive at the following expression
for the dynamically--screened interaction potential
in a nanoparticle:
\begin{eqnarray}\label{screen}
U(\omega;{\bf r},{\bf r}')
={u({\bf r}-{\bf r}')\over\epsilon(\omega)}
+ {e^2\over R}\sum_{LM}{4\pi\over 2L+1}
\frac{1}{\tilde{\epsilon}_L(\omega)}
\left({rr'\over R^2}\right)^L 
Y_{LM}(\hat{\bf r})Y^{\ast}_{LM}(\hat{\bf r}'),
\end{eqnarray}
with $\tilde{\epsilon}_L^{-1}(\omega)=
\epsilon_L^{-1}(\omega)-\epsilon^{-1}(\omega)$.
Equation (\ref{screen}), which is the main result of this section, 
represents a generalization of the plasmon
pole approximation to spherical particles. 
The two terms in the rhs describe two distinct
contributions. The first comes from 
 the usual  bulk-like screening of the Coulomb
potential. The second contribution describes a new 
effective {\em e--e} interaction induced by the {\em surface}: the
potential of an electron  inside the nanoparticle excites 
high--frequency surface  collective modes, which in turn act as
image charges that interact with the second electron.
It should be emphasized that, 
unlike in the case of the optical fields, the 
surface--induced dynamical screening
of the Coulomb potential
is  {\em  size--dependent}.

Note that the excitation energies of the surface collective
modes are lower than the bulk plasmon energy, also given by 
Eq.\ (\ref{omegaL}) but with $\epsilon_m=0$.  
This opens up new channels of quasiparticle scattering, considered
in the next section.  

\section{quasiparticle scattering via surface collective modes}
\label{sec:decay}

In this section we calculate the  rates of quasiparticle scattering
accompanied by the emission of  surface collective modes.
We start with the scattering of an electron in the conduction band.
In the first order in the surface--induced potential, given by the
second term in the rhs of Eq.\ (\ref{screen}), the corresponding
scattering rate can be obtained from  the Matsubara
self--energy\cite{mahan} 
\begin{eqnarray}\label{mselfc}
\Sigma_{\alpha}^{c}(i\omega)=
-\frac{1}{\beta}\sum_{i\omega'}\sum_{LM}\sum_{\alpha'}
\frac{4\pi e^2}{(2L+1)R^{2L+1}}
\frac{|M_{\alpha\alpha'}^{LM}|^2}{\tilde{\epsilon}_{L}(i\omega')}
G_{\alpha'}^{c}(i\omega'+i\omega),
\end{eqnarray}
where 
$G_{\alpha}^{c}=(i\omega-E_{\alpha}^c)^{-1}$ is the non-interacting
Green function of the conduction electron. Here the matrix elements
$M_{\alpha\alpha'}^{LM}$ are calculated with the one--electron 
wave functions $\psi_{\alpha}^c({\bf r})=R_{nl}(r)Y_{lm}(\hat{\bf r})$. 
Since $|\alpha\rangle$ and  $|\alpha'\rangle$ are the initial and
final states of the scattered electron,
the main contribution to  the $L$th term of the
angular momentum sum in Eq.\ (\ref{mselfc}) will come from electron
states with energy difference  
$E_{\alpha}-E_{\alpha'}\sim \omega_L$. Therefore, 
$M_{\alpha\alpha'}^{LM}$ can be expanded in terms of the small
parameter 
$E_0/|E_{\alpha}^c-E_{\alpha'}^c|\sim E_0/\omega_L$, where
$E_0=(2mR^2)^{-1}$ is the characteristic confinement energy.
The leading term can be
obtained by using the following procedure.\cite{kaw66,lus74}
We present $M_{\alpha\alpha'}^{LM}$ as
\begin{equation}
\label{melem}
M_{\alpha\alpha'}^{LM}
=\langle c,\alpha |r^LY_{LM}(\hat{\bf r})|c,\alpha'\rangle
=\frac{\langle c,\alpha |[H,[H,r^LY_{LM}(\hat{\bf r})]]|c,\alpha'\rangle}
{(E_{\alpha}^c-E_{\alpha'}^c)^2},
\end{equation}
where $H=H_0+V(r)$ is the Hamiltonian of an electron in a 
nanoparticle with confining potential $V(r)=V_0\theta(r-R)$. Since 
$[H,r^LY_{LM}(\hat{\bf r})]
=-\frac{1}{m}\nabla[r^LY_{LM}(\hat{\bf r})]\cdot\nabla$, the
numerator in Eq.\ (\ref{melem}) contains a term proportional to the
gradient of the confining potential, which peaks sharply at the 
surface. The corresponding 
contribution to the 
 matrix element describes the surface scattering of an
electron making 
the $L$--pole transition between the states $|c,\alpha\rangle$ and
$|c,\alpha'\rangle$, and gives the dominant  term of the
expansion. Thus, in the leading order in  
$|E_{\alpha}^c-E_{\alpha'}^c|^{-1}$, we obtain
\begin{equation}
\label{melem1}
M_{\alpha\alpha'}^{LM}=\frac{\langle c,\alpha |
\nabla [r^LY_{LM}(\hat{\bf r})]\cdot\nabla V(r)
|c,\alpha'\rangle}
{m(E_{\alpha}^c-E_{\alpha'}^c)^2}
=\frac{LR^{L+1}}{m(E_{\alpha}^c-E_{\alpha'}^c)^2}
V_0R_{nl}(R)R_{n'l'}(R)\varphi_{lm,l'm'}^{LM},
\end{equation}
with
$\varphi_{lm,l'm'}^{LM}=\int d\hat{\bf r}
Y_{lm}^{\ast}(\hat{\bf r})Y_{LM}(\hat{\bf r})Y_{l'm'}(\hat{\bf r})$.
Note that, for $L=1$, 
Eq.\ (\ref{melem1}) becomes exact.
For electron energies close to the Fermi level, $E_{nl}^c \sim E_F$,
the radial quantum  numbers are large, and the product 
$V_0 R_{nl}(R) R_{n'l'}(R)$ can be evaluated by using
semiclassical wave--functions.
In the limit $V_0\rightarrow \infty$, this product is given by 
\cite{kaw66} 
$2\sqrt{E_{nl}^cE_{n'l'}^c}/R^3$, where
$E_{nl}^c=\pi^2(n+l/2)^2E_0$ is the 
electron eigenenergy for large $n$.
Substituting this expression into  
Eq.\ (\ref{melem1}) and then into  Eq.\ (\ref{mselfc}),
we obtain
\begin{eqnarray}\label{mselfc1}
\Sigma_{\alpha}^{c}(i\omega)=
-\frac{1}{\beta}\sum_{i\omega'}\sum_{L}\sum_{n'l'}
C_{ll'}^{L}\,\frac{4\pi e^2 }{(2L+1)R}
\,\frac{E_{nl}^cE_{n'l'}^c}{(E_{nl}^c-E_{n'l'}^c)^4}
\,\frac{(4LE_0)^2}{\tilde{\epsilon}_{L}(i\omega')}
\,G_{\alpha'}^{c}(i\omega'+i\omega),
\end{eqnarray}
with
\begin{equation}
\label{C}
C_{ll'}^{L}=\sum_{M,m'}|\varphi_{lm,l'm'}^{LM}|^2
=\frac{(2L+1)(2l'+1)}{8\pi}\int_{-1}^{1}dxP_l(x)P_L(x)P_{l'}(x),
\end{equation}
where $P_l(x)$ are Legendre polynomials; we used properties
of the spherical harmonics in the derivation of Eq.\ (\ref{C}).
For  $E_{nl}^c\sim E_F$, the typical angular momenta are large, 
$l\sim k_{_F}R\gg 1$, and one can use the large--$l$ asymptotics of
$P_l$; for the low multipoles of interest, $L\ll l$, the
integral in Eq.\ (\ref{C}) can be
approximated by $\frac{2}{2l'+1}\delta_{ll'}$.
After performing the Matsubara summation, we
obtain for the imaginary part of the self--energy that 
determines the electron scattering rate 
\begin{equation}
\label{imselfc}
\mbox{Im}\Sigma_{\alpha}^{c}(\omega)=
-\frac{16e^2}{R}E_0^2\sum_L L^2\int dE \, g_l(E)
\frac{E E_{\alpha}^c}{(E_{\alpha}^c-E)^4}
\mbox{Im}\frac{N(E-\omega)+f(E)}
{\tilde{\epsilon}_{L}(E-\omega)},
\end{equation}
where  $N(E)$ is the Bose distribution and $g_l(E)$ is the
density of states of a conduction electron with angular momentum
$l$,
\begin{equation}
\label{gl}
g_l(E)=2\sum_{n}\delta(E_{nl}^c-E)\simeq
\frac{R}{\pi}\sqrt{\frac{2m}{E}},
\end{equation}
where we replaced the sum over $n$
by an integral (the factor of 2 accounts for spin).

Each term in the sum in the rhs of Eq.\ (\ref{imselfc}) represents
a channel of electron scattering mediated by a 
collective surface mode with angular momentum $L$. For low $L$,
the difference between the energies  of  modes
with successive values of $L$ is larger than their
widths, so that the different channels are well separated. 
Note that since all $\omega_L$ are smaller than the
frequency of the (undamped) bulk plasmon, one can replace 
$\tilde{\epsilon}_{L}(\omega)$ by  $\epsilon_{L}(\omega)$  in the
integrand of Eq.\ (\ref{imselfc}) for frequencies 
$\omega \sim \omega_{L}$.

Consider now the $L=1$ term in Eq.\ (\ref{imselfc}), which
describes the SP--mediated scattering channel. 
The main contribution to the integral comes from the SP pole in 
$\epsilon_1^{-1}(\omega)=3\epsilon_{s}^{-1}(\omega)$, where
$\epsilon_{s}(\omega)$ is the same as in 
Eq.~(\ref{epseff}). To estimate the scattering rate,
we approximate 
$\mbox{Im}\epsilon_{s}^{-1}(\omega)$ by  a
Lorentzian,
\begin{equation}
\label{lor}
\mbox{Im}\epsilon_{s}^{-1}(\omega)=
-\frac{\gamma_s\omega_p^2/\omega^3+\epsilon''_d(\omega)}
{[\epsilon'(\omega)+2\epsilon_m]^2+
[\gamma_s\omega_p^2/\omega^3+\epsilon''_d(\omega)]^2}
\simeq
-\frac{\omega_s^2}{\epsilon'_d(\omega_s)+2\epsilon_{m}}
\,
\frac{\omega_s\gamma}
{(\omega^2-\omega_s^2)^2+\omega_s^2\gamma^2},
\end{equation}
where 
$\omega_s\equiv \omega_1=\omega_p/\sqrt{\epsilon'_d(\omega_s)+2\epsilon_m}$
and $\gamma=\gamma_s+\omega_s\epsilon''_d(\omega_s)$ are the SP
frequency and width, respectively. For  typical widths
$\gamma\ll\omega_s$, the integral in Eq.\ (\ref{imselfc}) can be easily
evaluated, yielding
\begin{equation}
\label{imselfc1}
\mbox{Im}\Sigma_{\alpha}^{c}(\omega)=
-\frac{24e^2\omega_sE_0^2}{\epsilon'_d(\omega_s)+2\epsilon_m}
\frac{E_{\alpha}^c\sqrt{2m(\omega-\omega_s)}}
{(\omega-E_{\alpha}^c-\omega_s)^4}
[1-f(\omega-\omega_s)]. 
\end{equation}
Finally, using the relation
$e^2k_F[\epsilon'_d(\omega_s)+2\epsilon_m]^{-1}=3\pi\omega_s^2/8E_F$,
the SP--mediated scattering rate, 
$\gamma_e^s(E_{\alpha}^c)=-\mbox{Im}\Sigma_{\alpha}^{c}(E_{\alpha}^c)$,
takes the form  
\begin{equation}
\label{gammae}
\gamma_e^s(E)=9\pi\frac{E_0^2}{\omega_s}
\frac{E}{E_F}\left(\frac{E-\omega_s}{E_F}\right)^{1/2}
[1-f(E-\omega_s)].
\end{equation}
Recalling that $E_0=(2mR^2)^{-1}$, we see that the scattering rate of a 
conduction electron is {\em size--dependent}:
$\gamma_e^s\propto R^{-4}$. At $E=E_F+\omega_s$, the scattering
rate jumps to the value $9\pi(1+\omega_s/E_F)E_0^2/\omega_s$,
and then {\em increases} with energy as $E^{3/2}$ 
(for $\omega_s\ll E_F$). This should be 
contrasted with the usual (bulk) plasmon--mediated  scattering,
originating from the first term in Eq.\ (\ref{screen}), with the
rate decreasing as $E^{-1/2}$ above the onset.\cite{mahan} 
To estimate the size at which $\gamma_e^s$ becomes important, we
should compare it with the Fermi liquid  
{\em e--e} scattering rate,\cite{pines} 
$\gamma_e(E)=\frac{\pi^2q_{_{TF}}}{16k_{_F}}\frac{(E-E_F)^2}{E_F}$.
For energies $E\sim E_F+\omega_s$, the two rates
become comparable for
\begin{equation}
\label{size}
(k_{_F}R)^2\simeq
12\frac{E_F}{\omega_s}\left(1+\frac{E_F}{\omega_s}\right)^{1/2}
\left(\frac{k_{_F}}{\pi q_{_{TF}}}\right)^{1/2}.
\end{equation}
In the case of a Cu
nanoparticle with $\omega_s\simeq 2.2$ eV, we obtain 
$k_{_F}R\simeq 8$, which corresponds  to the radius $R\simeq 3$ nm. 
At the same time, in this energy range, the width $\gamma_e^s$ exceeds
the mean level spacing $\delta$, so that the energy spectrum is
still continuous.
The strong size dependence of $\gamma_e^s$ indicates that, although
$\gamma_e^s$
increases with energy slower than $\gamma_e$, 
the SP--mediated scattering should dominate for nanometer--sized
particles. Note that the size and energy dependences of scattering
in different channels are similar. Therefore,  
the total scattering rate as a function of energy will represent
a series of steps at the collective excitation energies
$E=\omega_L<\omega_p$  on top of a smooth energy increase. 
We expect that this effect could be observed experimentally in
time--resolved two--photon photoemission measurements of
size--selected cluster beams.\cite{petek}

We now turn to the interband processes in noble metal particles 
and consider the scattering of a $d$--hole into the conduction band.
From now on we restrict ourselves to the scattering via the dipole
channel, mediated by the SP. The corresponding surface--induced
potential, given by the  $L=1$ term in Eq.\ (\ref{screen}), has the form 
\begin{eqnarray}
\label{spscreen}
U_s(\omega;{\bf r},{\bf r}')=
{3e^2\over R}{{\bf r}\cdot {\bf r}'\over R^2}
{1\over \epsilon_{s}(\omega)}.
\end{eqnarray}
With this potential, the $d$--hole self--energy is given by
\begin{eqnarray}\label{mselfd}
\Sigma_{\alpha}^{d}(i\omega)=
-{3e^2\over R^3}\sum_{\alpha'}|{\bf d}_{\alpha\alpha'}|^2
{1\over \beta}\sum_{i\omega'}
\frac{G_{\alpha'}^{c}(i\omega'+i\omega)}{\epsilon_{s}(i\omega')},
\end{eqnarray}
where
${\bf d}_{\alpha\alpha'}=\langle c,\alpha |{\bf r}|d,\alpha'\rangle=
\langle c,\alpha |{\bf p}|d,\alpha'\rangle/im(E^c_{\alpha}-E^d_{\alpha'})$ 
is the interband transition matrix element.
Since the final state energies in the conduction band are high 
(in the case of interest here, they are close to
the Fermi level), the matrix element can be 
approximated by the bulk--like expression
$\langle c,\alpha |{\bf p}|d,\alpha'\rangle
=\delta_{\alpha\alpha'}\langle c|{\bf p}|d\rangle
\equiv \delta_{\alpha\alpha'}\mu$, 
the corrections due 
to  surface scattering 
being suppressed by a factor of $(k_{_F}R)^{-1}\ll 1$. 
After performing the 
frequency summation, we obtain for Im$\Sigma_{\alpha}^{d}$

\begin{eqnarray}\label{imselfd}
{\rm Im}\Sigma_{\alpha}^{d}(\omega)
=-{9e^2 \mu^2\over m^2(E^{cd}_{\alpha})^2R^3}
\mbox{Im}\frac{N(E^c_{\alpha}-\omega)+f(E^c_{\alpha})}
{\epsilon_{s}(E^{c}_{\alpha}-\omega)},
\end{eqnarray} 
with $E^{cd}_{\alpha}=E^{c}_{\alpha}-E^{d}_{\alpha}$.
We see that the scattering rate of a $d$-hole with energy $E^{d}_{\alpha}$, 
$\gamma_{h}^s(E^{d}_{\alpha})=\mbox{Im}\Sigma_{\alpha}^{d}(E^{d}_{\alpha})$,
has a strong $R^{-3}$ dependence on the nanoparticle size, which is,
however, different from that of the intraband scattering, 
Eq.\ (\ref{gammae}). 

The important difference between the interband
and the intraband SP--mediated scattering rates 
lies in their energy dependence.
Since the surface--induced  potential, Eq.\ (\ref{spscreen}),
allows for only vertical (dipole) interband single--particle
excitations, the phase space for the scattering of a
$d$--hole with energy $E^{d}_{\alpha}$ is
restricted to a single final state in the conduction band
with energy $E^{c}_{\alpha}$. As a result, 
the $d$--hole scattering rate,
$\gamma_{h}^s(E^{d}_{\alpha})$, exhibits a {\em peak} 
as the difference between the energies of final and initial states,  
$E^{cd}_{\alpha}=E^{c}_{\alpha}-E^{d}_{\alpha}$,
approaches the SP frequency $\omega_{s}$ 
[see Eq.\ (\ref{imselfd})].  
In contrast, the energy dependence of $\gamma_e^s$ is smooth due
the larger phase space available for scattering
in the conduction band. 
This leads to the additional integral over final state energies in
Eq.\ (\ref{imselfc}), which smears out the SP resonant enhancement
of the intraband scattering. 

As we show in the next section, 
the fact that the  scattering rate of a $d$--hole is 
dominated by the SP resonance, affects strongly the nonlinear
optical dynamics in small nanoparticles. 
This is the case, in particular, when the SP frequency,
$\omega_{s}$, is close to the onset of interband transitions,
$\Delta$, as, e.g., in Cu and Au 
nanoparticles.\cite{kreibig,big95,per97,kla98}
Consider an  {\em e--h} pair with excitation energy
$\omega$ close to $\Delta$. As we discussed, the $d$--hole can
scatter into the conduction band by emitting a SP. According to
Eq.~(\ref{imselfd}), for $\omega\sim\omega_{s}$, 
this process will be resonantly enhanced. 
At the same time, the electron can 
scatter in the conduction band  via the usual
two--quasiparticle process. For $\omega\sim\Delta$, the electron energy
is close to $E_F$, and its scattering rate is estimated as\cite{petek}
$\gamma_{e}\sim 10^{-2}$ eV. Using the bulk value of $\mu$,
$2\mu^2/m\sim 1$ eV near the L-point,\cite{eir62} we find that 
$\gamma_{h}^s$ exceeds $\gamma_{e}$ for 
$R\lesssim 2.5$\ nm. In fact, one would expect
that, in nanoparticles, $\mu$ is larger than in the bulk due to the
localization of the conduction electron  wave--functions.\cite{kreibig}

\section{Surface plasmon nonlinear optical dynamics}
\label{sec:optics}

In this section, we study the effect of the SP--mediated 
{\em interband} scattering on the nonlinear optical dynamics
in noble metal nanoparticles. 
When the hot electron
distribution has already thermalized and the electron gas is cooling
to the lattice, the transient response of a nanoparticle can be
described by the time--dependent absorption coefficient
$\alpha(\omega,t)$, given by  Eq.\ (\ref{absor}) with time--dependent
temperature.\cite{perakis} In noble--metal particles, the
temperature dependence of $\alpha$ originates
from two different sources. First is the phonon--induced
correction to $\gamma_s$, which is proportional to the {\em lattice}
temperature $T_l(t)$. As mentioned in the Introduction, for small
nanoparticles this effect is relatively weak. Second, near the onset
of the interband transitions, $\Delta$, the absorption coefficient
depends on the {\em electron} temperature $T(t)$
via the interband dielectric function 
$\epsilon_d(\omega)$ [see Eqs.\ (\ref{absor}) and (\ref{epseff})].
In fact, in Cu or Au nanoparticles, 
$\omega_s$ can be tuned close to $\Delta$, so  
the SP damping by {\em interband} {\em e--h}
excitations leads to an additional broadening of the absorption
peak.\cite{kreibig} In this case, the temperature dependence of 
$\epsilon_d(\omega)$ dominates the pump--probe 
dynamics. Below we show that, near the SP resonance, both the
temperature and frequency dependence of
$\epsilon_d(\omega)=1+4\pi\chi_d(\omega)$ are strongly affected by
the SP--mediated interband scattering.

For non-interacting electrons, the interband susceptibility, 
$\chi_d(i\omega)=\tilde{\chi}_d(i\omega)+\tilde{\chi}_d(-i\omega)$,
has the standard form\cite{mahan}
\begin{equation}\label{susc}
\tilde{\chi}_d(i\omega)=
-\sum_{\alpha}{e^2\mu^2\over m^2(E^{cd}_{\alpha})^2}
{1\over \beta}\sum_{i\omega'}
G_{\alpha}^{d}(i\omega')
G_{\alpha}^{c}(i\omega'+i\omega),
\end{equation}
where $G_{\alpha}^{d}(i\omega')$ is the Green function of a
$d$--electron. Since the $d$-band is fully occupied,
the only allowed SP--mediated interband scattering 
is that of the  $d$--hole. We assume here, for simplicity, a
dispersionless $d$--band with energy $E^d$.
Substituting  $G_{\alpha}^{d}(i\omega')= 
[i\omega'-E^{d}+E_F-\Sigma_{\alpha}^{d}(i\omega')]^{-1}$,
with $\Sigma_{\alpha}^{d}(i\omega)$ given by Eq.~(\ref{mselfd}),
and  performing the frequency summation, we  obtain
\begin{equation}\label{interband}
\tilde{\chi}_d(\omega)=
{e^2 \mu^2\over m^2}\int {dE^c\, g(E^c)\over (E^{cd})^2}
{f(E^c)-1\over \omega-E^{cd}+i\gamma_h^s(\omega,E^c)},
\end{equation}
where $g(E^c)$ is the density of states of
conduction electrons. Here
$\gamma_h^s(\omega,E^c)={\rm Im}\Sigma^{d}(E^c-\omega)$ is the
scattering rate of a $d$-hole with energy $E^c-\omega$, for which we
obtain from  Eq.\ (\ref{imselfd}),
\begin{equation}\label{gamhole}
\gamma_h^s(\omega,E^c)
=-{9e^2 \mu^2\over m^2(E^{cd})^2R^3}
f(E^c)\mbox{Im}{1\over\epsilon_{s}(\omega)},
\end{equation}
where we neglected $N(\omega)$ for frequencies
$\omega\sim\omega_s\gg k_BT$.
Remarkably, $\gamma_h^s(\omega,E^c)$ exhibits a 
sharp peak  as a function of the {\em frequency of the 
probe optical field}. The reason 
for this is that the scattering rate of a $d$--hole with energy $E$
depends explicitly on the  {\em difference} between 
the final and initial states, $E^c-E$,
as discussed in the previous section: therefore, for a $d$--hole
with energy $E=E^c-\omega$, the dependence on the final 
state energy, $E^c$, cancels out in $\epsilon_s(E^c-E)$ 
[see Eq.\ (\ref{imselfd})].
This implies that the optically--excited 
$d$--hole experiences a {\em resonant scattering} into the
conduction band as the probe frequency 
$\omega$ approaches the SP frequency. 
It is important to note that 
$\gamma_h^s(\omega,E^c)$ is, in fact, proportional to the 
absorption coefficient $\alpha(\omega)$ [see Eq.\ (\ref{absor})].
Therefore, the calculation of the absorption 
spectrum is a {\em self--consistent} problem defined 
by Eqs.\ (\ref{absor}), (\ref{epseff}), (\ref{interband}), and
(\ref{gamhole}).

It should be emphasized that  the effect of $\gamma_h^s$ on
$\epsilon_d''(\omega)$  {\em increases with temperature}.
Indeed, the Fermi function in the rhs of Eq.\ (\ref{gamhole}) implies that 
$\gamma_h^s$ is small unless $E^c-E_F \lesssim k_B T$. Since the main
contribution to $\tilde{\chi}''_d(\omega)$
comes from energies $E^c-E_F \sim \omega-\Delta$, the
$d$--hole scattering becomes  efficient for
electron temperatures $k_B T\gtrsim \omega_{s}-\Delta$. 
As a result, near the SP resonance, the time evolution of the
differential absorption, governed by the temperature dependence of $\alpha$, 
becomes strongly size--dependent, as we illustrate in the rest of
this section. 

In the numerical calculations below, we adopt the parameters of the
experiment of Ref. \onlinecite{big95}, which was performed on 
$R\simeq 2.5$ nm Cu nanoparticles with SP frequency,   
$\omega_{s}\simeq 2.22$ eV, slightly above the onset of the
interband transitions, $\Delta\simeq 2.18$ eV.
In order to describe  the time--evolution
of the differential absorption spectra, we first need to
determine the time--dependence of the electron temperature, $T(t)$,
due to the relaxation of the electron gas to the lattice. 
For this, we employ a simple two--temperature model,
defined by heat equations for 
$T(t)$ and the lattice temperature $T_l(t)$:
\begin{eqnarray}
\label{TT}
C(T)\frac{\partial T}{\partial t}&=&-G(T-T_l),
\nonumber\\ 
C_l\frac{\partial T_l}{\partial t}&=&G(T-T_l),
\end{eqnarray}
where $C(T)=\Gamma T$ and $C_l$ are the electron and lattice heat
capacities, respectively, and $G$ is 
the electron--phonon coupling.\cite{eas86}
The parameter values used here were $G=3.5\times 10^{16}$ Wm$^{-3}$K$^{-1}$, 
$\Gamma=70$ Jm$^{-3}$K$^{-2}$, and $C_l=3.5$ Jm$^{-3}$K$^{-1}$.
The values of $\gamma_s$ and $\mu$ were extracted from the fit to
the linear absorption spectrum, and the initial condition 
for Eq.\ (\ref{TT}) was taken as $T_0=800$ K, the estimated
pump--induced hot electron temperature.\cite{big95}
We then self--consistently calculated the time--dependent absorption
coefficient $\alpha(\omega,t)$, and the differential
transmission is proportional to
$\alpha_r(\omega)-\alpha(\omega,t)$, where $\alpha_r(\omega)$ was
calculated at the room temperature.

In Fig.\ 1 we plot the calculated differential transmission spectra
for different nanoparticle sizes. Fig.\ 1(a) shows the spectra
at several time delays for $R=5.0$\ nm; in this case, the
SP--mediated {\em d}--hole scattering has no significant effect.  
Note that it is necessary to  include the intraband
{\em e--e} scattering in order to reproduce  
the differential transmission 
lineshape observed in the experiment.\cite{big95} 
For optically excited electron energy close to $E_F$, this  can be 
achieved  by adding the {\em e--e} scattering rate\cite{pines} 
$\gamma_e(E^c)\propto [1-f(E^c)][(E^c-E_F)^2+(\pi k_B T)^2]$ to
$\gamma_h^s$ in Eq.\ (\ref{interband}). The difference in 
$\gamma_e(E^c)$ for $E^c$ below and above $E_F$ leads to a
lineshape similar to that expected from the combination of
red--shift and broadening.

In Figs.\ 1(b)\ and\ (c) we show the differential transmission spectra 
with decreasing nanoparticle size.  
For $R=2.5$ nm, the apparent red--shift is reduced 
[see Fig.\ 2(b)]. This change can be explained  as follows.  
Since here $\omega_{s}\sim\Delta$, the SP is damped by the interband
excitations for $\omega>\omega_{s}$, so that the absorption peak is 
{\em asymmetric}. The $d$--hole scattering with the SP
enhances the damping; however, since the $\omega$--dependence of 
$\gamma_h^s$ follows that of $\alpha$, this effect is larger above
the resonance. On the other hand, the efficiency of scattering
increases with temperature, as discussed above. Therefore, for short
time delays, the increase in the absorption is relatively larger for
$\omega>\omega_{s}$.
With decreasing size, the strength  of this effect increases,
leading to an apparent blue--shift [see Fig.\ 2(c)].  
Such a strong change in the absorption dynamics
originates from the $R^{-3}$ dependence of the $d$--hole scattering
rate; reducing the size by the factor of two results in an
enhancement of $\gamma_h^s$  by an order of magnitude.

In   Fig.\ 2 we show the time evolution of the differential
transmission at several frequencies close to $\omega_s$. It can be
seen that the relaxation is slowest at the SP resonance;
this characterizes the robustness of the collective mode, which
determines the peak position, versus the single--particle
excitations, which determine the resonance width. 
For larger sizes, at which $\gamma_h^s$ is small, the change in  
the differential transmission decay rate with {\em frequency} 
is smoother above the resonance [see Fig. 2(a)]. 
This stems from the asymmetric lineshape of the
absorption peak, mentioned above: the  absorption is
larger for $\omega>\omega_{s}$, so that its {\em relative} change 
with temperature is weaker.
For smaller nanoparticle size, the decay rates become similar above
and below  $\omega_s$  [see Fig.\ 2(b)]. This change in the
frequency dependence is related to the stronger SP damping for
$\omega>\omega_{s}$ due to the $d$--hole scattering, as discussed
above. Since this additional damping is reduced with decreasing
temperature, the relaxation is faster above the
resonance, compensating the relatively weaker change in the
absorption. This rather ``nonlinear'' relation between the
time--evolution of the pump--probe signal and that of the temperature,
becomes even stronger for smaller sizes [see Fig.\ 2(c)]. In this
case, the frequency dependence of the differential transmission
decay  below and above $\omega_s$ is reversed. Note, that a
frequency dependence consistent with our calculations presented 
in Fig.\ 2(b) was, in fact, observed in the experiment of
Ref.\onlinecite{big95}. At the same time, the changes in the linear
absorption spectrum are relatively small.

\section{Conclusions}

To summarize, we have examined theoretically the role of
size--dependent correlations in the electron relaxation in small 
metal particles. We identified a new mechanism of quasiparticle
scattering, mediated by collective surface excitations, which
originates from the surface--induced dynamical
screening of the {\em e--e} interactions. The behavior of the
corresponding scattering rates with varying energy and temperature
differs substantially from that in the bulk metal. In particular, in
noble metal particles, the energy dependence of the $d$--hole
scattering rate was found similar to that of the absorption
coefficient. This led us to a self--consistent scheme for the
calculation of the absorption spectrum near the surface plasmon 
resonance.

An important aspect of the SP--mediated scattering is its strong
dependence on size. Our estimates show that it
becomes comparable to the usual Fermi--liquid scattering in
nanometer--sized particles. This size regime is, in fact,
intermediate between ``classical'' particles with sizes larger
than 10 nm, where the bulk--like behavior dominates, and
very small clusters with only dozens of atoms, where the metallic
properties are completely lost. Although the static properties
of nanometer--sized particles are also size--dependent,
the deviations from their bulk values do not change the qualitative
features of the electron {\em dynamics}. In contrast, the
size--dependent {\em many--body} effects, studied here, {\em do}
affect the dynamics in a significant way during time scales comparable
to the relaxation times. 
As we have shown, the SP--mediated interband scattering
reveals itself in the transient pump--probe spectra. In particular,
as the nanoparticle size decreases, 
the calculated time--resolved  differential absorption 
develops a characteristic lineshape corresponding to a resonance
blue--shift. At the same time, near the SP resonance, the 
scattering leads to a significant change in the frequency dependence
of the relaxation time of the pump--probe signal, consistent with recent 
experiments. 
These results indicate the need for a systematic
experimental studies of   the size--dependence of the transient
nonlinear optical  response, as we approach the transition 
from boundary--constrained nanoparticles to molecular 
clusters.

The authors thank J.--Y. Bigot for valuable discussions.
This work was supported by NSF CAREER award ECS-9703453, and, in
part, by ONR Grant N00014-96-1-1042  and by Hitachi Ltd.

\begin{figure}
\caption{
Calculated differential transmission spectra at positive time
delays for nanoparticles with  
(a) $R=5$ nm, (b) $R=2.5$ nm, and (c) $R=1.2$ nm.
}
\end{figure}

\begin{figure}
\caption{
Temporal evolution of the differential transmission 
at frequencies close the SP resonance for nanoparticles
with  
(a) $R=5$ nm, (b) $R=2.5$ nm, and (c) $R=1.2$ nm.
}
\end{figure}

\clearpage

\epsfxsize=6.0in
\epsffile{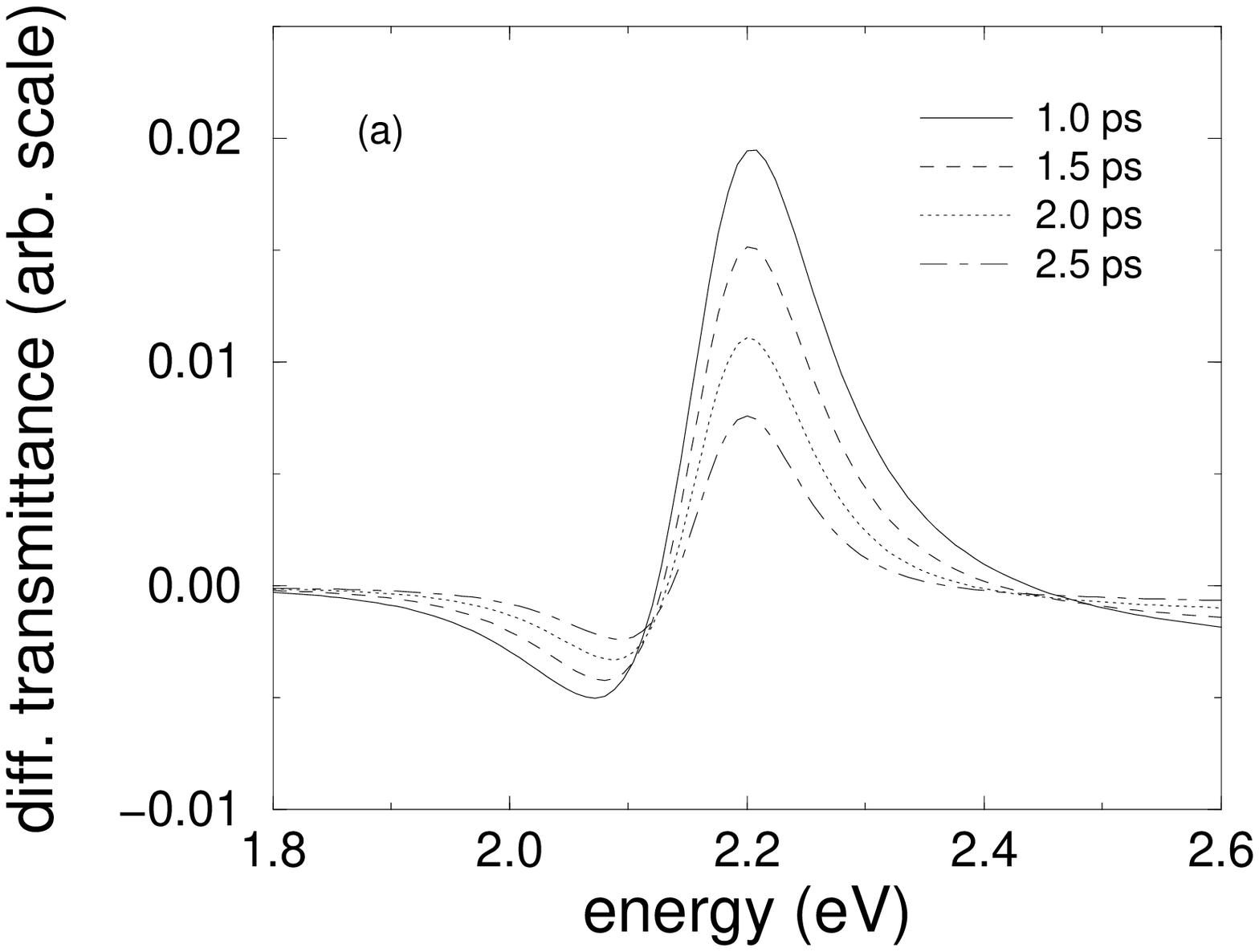}
\vspace{80mm}
\centerline{FIG. 1}
\epsfxsize=6.0in
\epsffile{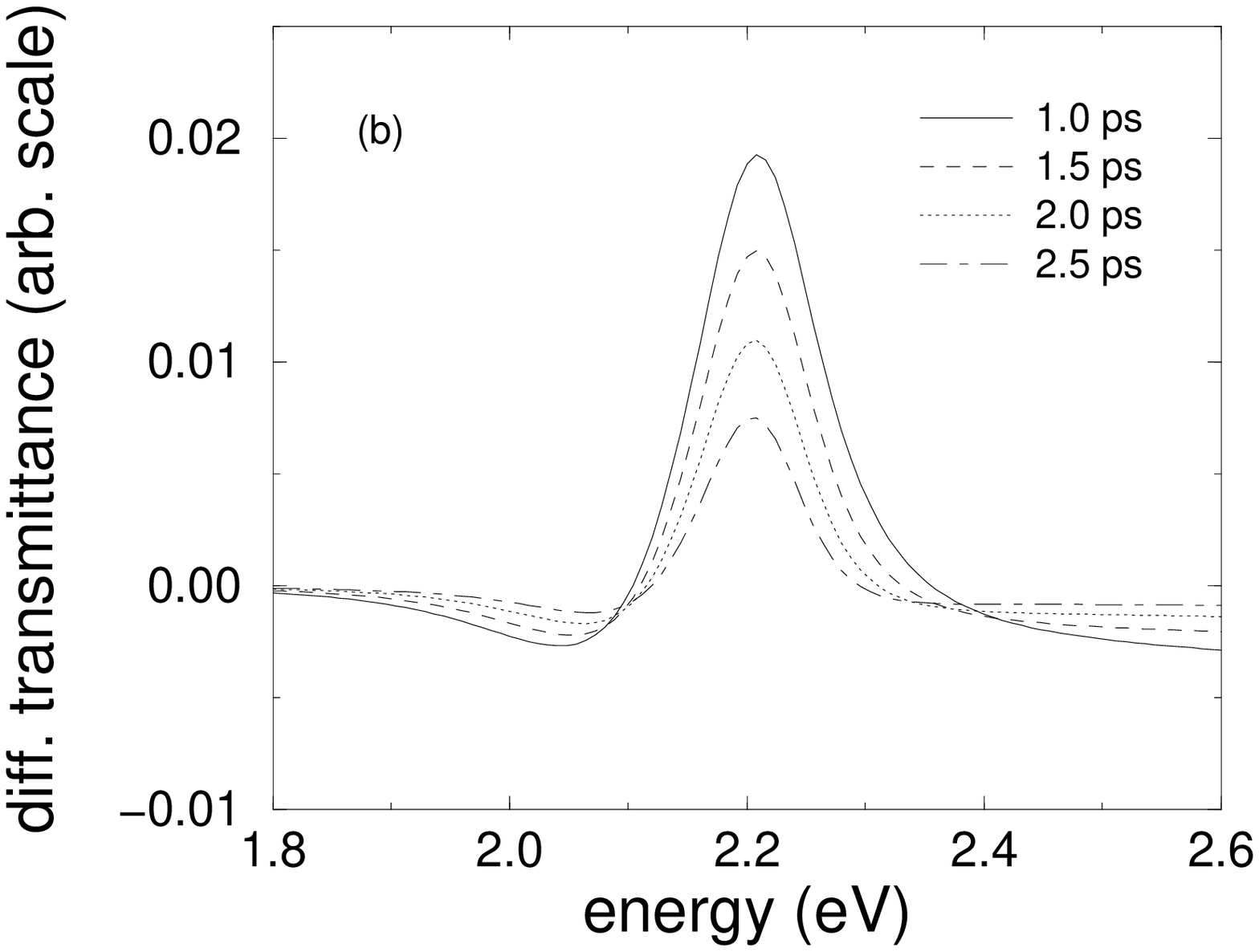}
\vspace{80mm}
\centerline{FIG. 1}
\epsfxsize=6.0in
\epsffile{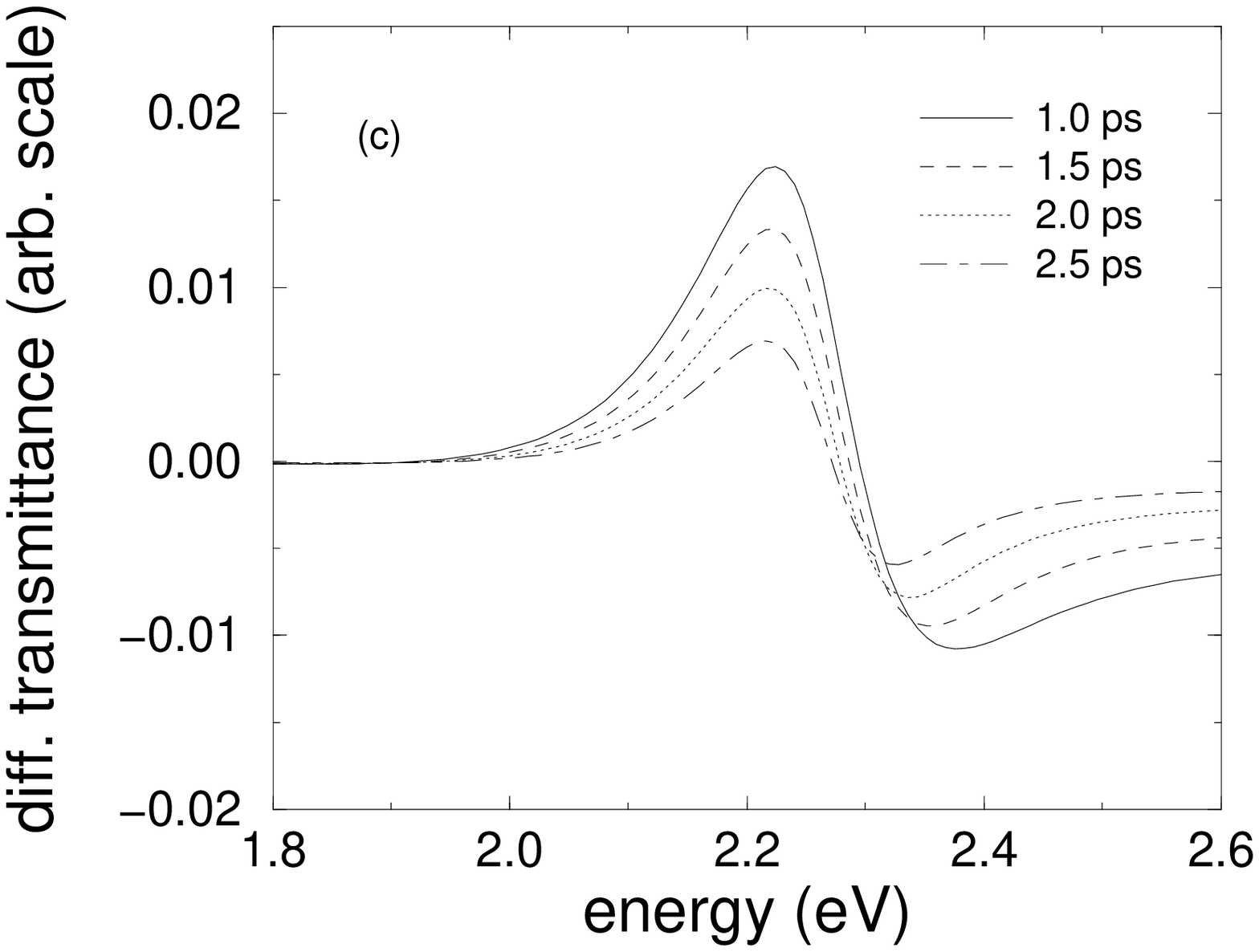}
\vspace{80mm}
\centerline{FIG. 1}
\epsfxsize=6.0in
\epsffile{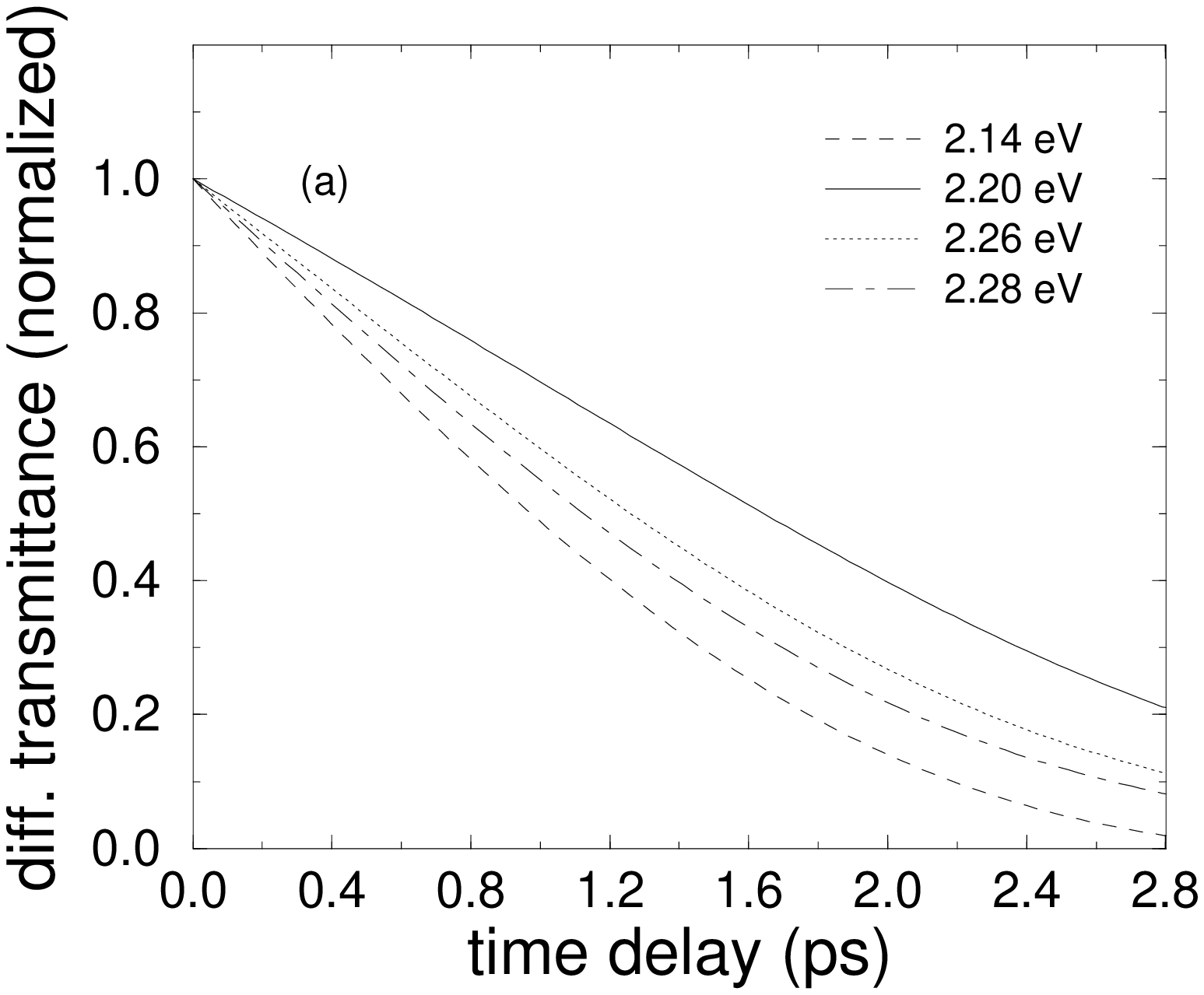}
\vspace{80mm}
\centerline{FIG. 2}
\epsfxsize=6.0in
\epsffile{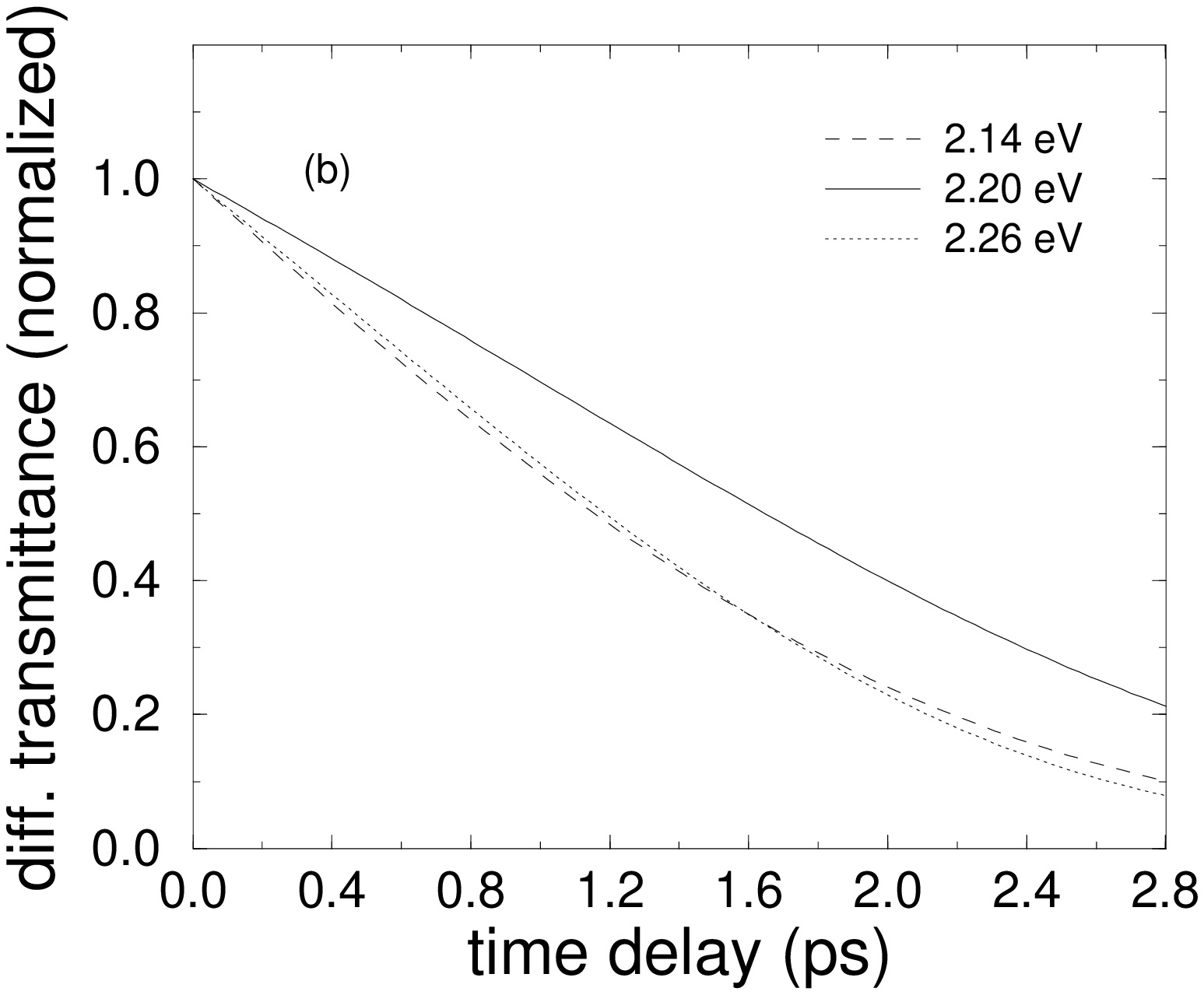}
\vspace{80mm}
\centerline{FIG. 2}
\epsfxsize=6.0in
\epsffile{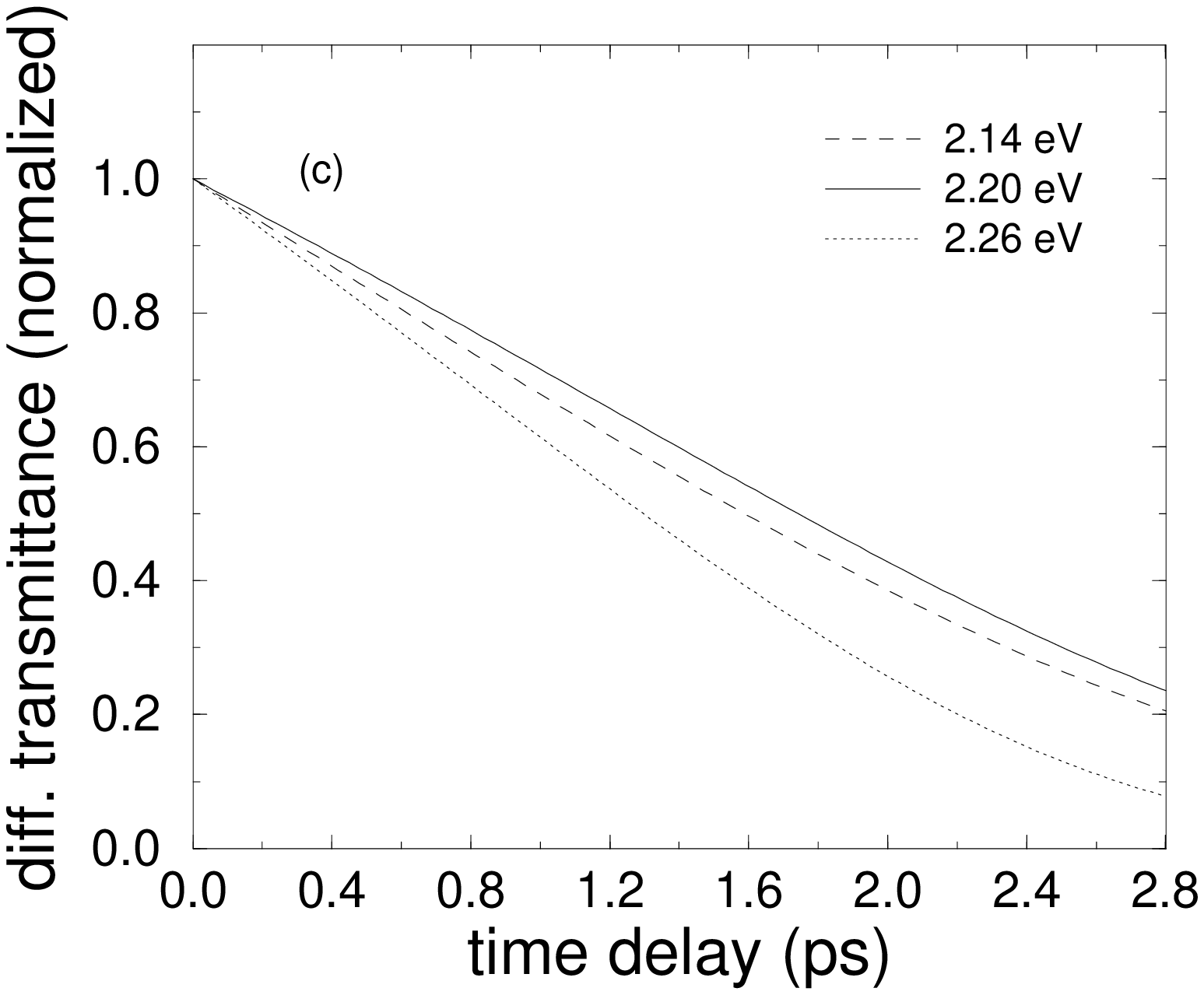}
\vspace{80mm}
\centerline{FIG. 2}

\end{document}